\documentclass[iop]{emulateapj}

\usepackage{amsmath}

\begin{document}

\title{Insight Into ``Changing-Look'' AGN Mrk 1018 from the Fe K$\alpha$ Line: The Reprocessing Gas Has Yet to Fully Respond to the Fading of the AGN}

\author{Stephanie M. LaMassa$^{1}$, Tahir Yaqoob$^2$, Roy Kilgard$^3$}
\affil{
$^1$Space Telescope Science Institute, 3700 San Martin Drive, Baltimore, MD, 21218, USA;
$^2$Department of Physics, University of Maryland Baltimore County, 1000 Hilltop Circle, Baltimore, MD 21250, USA;
$^3$Astronomy Department, Wesleyan University, Middletown, CT 06459, USA
}
\begin{abstract}
  Mrk 1018 is a ``changing-look'' AGN whose optical spectrum transitioned from a Type 1.9 to Type 1 between 1979 and 1984 and back to a Type 1.9 in 2015. This latest transition was accompanied by a decrease in X-ray flux. We analyze the {\it Chandra} spectra from 2010 and 2016 and {\it NuSTAR} spectra from 2016, carefully treating pile-up in the {\it Chandra} spectrum from 2010 and self-consistently modeling absorption, reflection, and Fe K$\alpha$ line emission in the X-ray spectra from 2016. We demonstrate that while the 2-10 keV X-ray flux decreased by an order of magnitude (1.46$^{+0.10}_{-0.13} \times 10^{-11}$ erg s$^{-1}$ cm$^{-2}$ to 1.31$^{+0.09}_{-0.04} \times 10^{-12}$ erg s$^{-1}$ cm$^{-2}$), the Fe K$\alpha$ equivalent width (EW) increased from 0.18$^{+0.17}_{-0.12}$ to 0.61$^{+0.27}_{-0.25}$ keV, due to a depressed AGN continuum. We jointly fit the {\it Chandra} and {\it NuSTAR} spectra from 2016 using the physically-motivated MYTorus model, finding that the torus orientation is consistent with a face-on geometry, and lines of sight intersecting the torus are ruled out. While we measure no line-of-sight absorption, we measure a column density of $N_{\rm H}$ = 5.38$^{+14}_{-4.0} \times 10^{22}$ cm$^{-2}$ for gas out of the line-of-sight which reprocesses the X-ray emission. We find a high relative normalization between the Compton scattered emission and transmitted continuum, indicative of time lags between the primary X-ray source and reprocessing gas. We predict that the Fe K$\alpha$ line will respond to the decrease in AGN flux, which would manifest as a decrease in the Fe K$\alpha$ EW.

\end{abstract}

\section{Introduction}
Active galactic nuclei (AGN) are powered by accretion onto a supermassive black hole, liberating energy that shapes their surroundings. In a hot corona near the accretion disk, optical and ultraviolet photons are Compton upscattered to X-ray energies. Gas near the black hole, photoionized by accretion disk photons, orbits rapidly, imprinting Doppler broadened lines on the optical spectrum. Slower moving gas, hundredes to thousands of parsecs away,  can still be energized by the accretion disk which produces prominent narrow emission lines in optical spectra \citep{bpt}. In some AGN, both broad and narrow spectral emission lines are observed (Type 1), while only narrow emission lines are observed in others (Type 2). According to the AGN unification model \citep{antonucci,urry}, a parsec scale torus of dust and gas enshrouds the central engine such that the line-of-sight to this system determines whether or not broad lines are visible. This obscuring material can have a sufficiently high column density (i.e., $N_{\rm H} > 10^{22}$ cm$^{-2}$) to affect even powerful X-ray emission.

However, the unified model fails to explain the spectral states of some AGN. So-called ``changing-look'' AGN are sources that have transitioned optical spectral type, from Type 1 to intermediate types (Type 1.8 and 1.9) to Type 2, or vice versa \citep{tohline,penston,tran}. Changes in the X-ray emission have been observed in other AGN \citep{risaliti2009,risaliti2011,ricci}, sometimes accompanied by optical spectroscopic transitions \citep{shappee}. Though some of these sources can be explained by the passage of occulting clouds, in concordance with the unified model \citep[e.g.,][]{goodrich,storchi-bergmann,risaliti2009,risaliti2011}, others are more likely triggered by a change in the ionizing power emanating from the accretion disk \citep{lamassa2015,merloni,ruan,runnoe,macleod}.

Mrk 1018 is a particuarly interesting changing-look AGN as it is one of only 4 sources known to have transitioned across the spectral sequence twice \citep{aretxaga,denney,macleod}. \citet{cohen} reported that Mrk 1018 changed from a Type 1.9 to Type 1 between 1979 and 1984. More recently, \citet{mcelroy} discovered that Mrk 1018 returned to the Type 1.9 state, with the broad lines disappearing between 2009 and 2015. Analysis of the optical spectrum suggested that a decrease in the accretion rate was responsible for the spectral transition. The X-ray properties of this source were reported in \citet{husemann} where they demonstrated that the observed 2-10 keV flux decreased by a factor of 7.5 between a {\it Chandra} observation in 2010 and follow-up {\it Chandra} and {\it NuSTAR} observations in 2016, consistent with the time scale in which the optical continuum diminished and broad lines disappeared. They claimed that the {\it Chandra} and {\it NuSTAR} spectra show no signs of obscuration.

\citet{husemann} fitted the {\it Chandra} and {\it NuSTAR} spectra with an absorbed powerlaw, and though noted that Fe K$\alpha$ fluorescent line emission is present in the X-ray spectra from 2016, did not fit this feature. Fe K$\alpha$ emission results when X-ray photons with energies above the K-edge threshold (7.124 keV) are absorbed, liberating a K-shell electron which is then filled by the decay of an electron from an upper level, resulting in the emission of a fluorescent line photon at 6.4 keV. Fe K$\alpha$ emission at 6.4 keV then signifies the reprocessing of X-ray emission by nearby gas, providing clues to the geometry and column density of the reprocessing medium.

Here we revisit the analysis of the {\it Chandra} and {\it NuSTAR} spectra of Mrk 1018 since column densities derived via absorbed single power law fits to X-ray spectra can be unreliable \citep{turnera,turnerb,panessa,lamassa2009,lamassa2011}. We first examine the {\it Chandra} spectra from the 2010 and 2016, taking care to account for the effects of pile-up (a phenomenon where two X-ray photons are read as a single event in bright sources) in the {\it Chandra} spectrum from 2010. From this exercise we measure the Fe K$\alpha$ equivalent width (EW) from both epochs of X-ray observations, since an increase in the EW may signal an increase in the line-of-sight obscuration \citep{krolik,ghisellini,turnerb,levenson}: while the Fe K$\alpha$ line photons form within the obscuring medium, the transmitted continuum against which the line is measured is suppressed. However, time delays between the continuum and the gas reprocessing the X-ray emission can also conspire to boost the Fe K$\alpha$ EW. We then self-consistently model the transmitted, reflected, and Fe K$\alpha$ emission from the {\it Chandra} and {\it NuSTAR} spectra in 2016 using the physically motivated MYTorus model \citep{mytorus}. From this modeling, we measure the column density of the gas responsible for reprocessing the X-ray emission and gain insight into the physical processes that drive the observed change in X-ray flux.

\section{Data Analysis}\label{da}

\subsection{\textit{Chandra} Data Reduction}
Mrk 1018 was first observed on 2010 November 27 for 23 ks (PI: Mushotzky; ObsID: 12868) with a 1/8 subarray (frame time = 0.4 s) on ACIS-S. Pursuant to the discovery of the optical spectroscopic transition, it was re-observed under a Director Discretionary Time (DDT) request on 2016 February 6 for 27 ks (PI: Tremblay; ObsID: 18789), with the same set-up as the observation from 2010. We reduced both spectra with \textsc{CIAO v4.8}, with \textsc{CALDB v. 4.7.2}. We used \textsc{CIAO} task \textsc{chandra\_repro} to produce a filtered level 2 events file with the latest calibration files.

The {\it Chandra} spectra from both epochs of observation were extracted with the \textsc{CIAO} routine \textsc{specextract}. Due to the high count-rate from Mrk 1018 in 2010 ($\sim$2.4 counts s$^{-1}$), the first spectrum was heavily affected by pile-up which occurs when two or more photons are recorded as a single event by the CCD detector. This results in under-sampling of the point spread function (PSF) and flattening of the spectrum. For piled-up sources, it is recommended that the spectrum be extracted from a 2$^{\prime\prime}$ radius, corresponding to 4 ACIS pixels \citep{davis}. As the latter observation was not piled, we extracted the spectrum with a 3$^{\prime\prime}$ radius to encompass all the flux from the source. For both observations, we extracted the background spectrum from an annulus around the source with an inner radius of 5$^{\prime\prime}$ and outer radius 15$^{\prime\prime}$. The data were grouped with a minimum of 15 counts per bin.

\subsection{\textit{NuSTAR} Data Reduction}
           {\it NuSTAR} is a focusing hard X-ray telescope with two co-aligned focal plane modules, FPMA and FPMB, providing coverage from 3 to 79 keV \citep{harrison}. Mrk 1018 was observed by {\it NuSTAR} in 2016 February 10 for 21 ks as part of the Extended Groth Strip (EGS) extragalactic survey. Using \textsc{NuSTARDAS} v1.6.0, we ran \textsc{nupipeline} to produce filtered events files. From these files, we extracted the source spectrum from a 45$^{\prime\prime}$ radius, using the \textsc{nuproduct} task; the background was extracted from a source-free region on the detector. We extracted spectra separately from FPMA and FPMB, and grouped by a mininum of 15 counts per bin.

In the spectral modeling below, all errors are reported at the 90\% confidence level.

\section{Results}

\subsection{Accounting for Pile-up in the Chandra Spectrum from 2010}
\begin{figure*}
  \centering
  \includegraphics[scale=0.35]{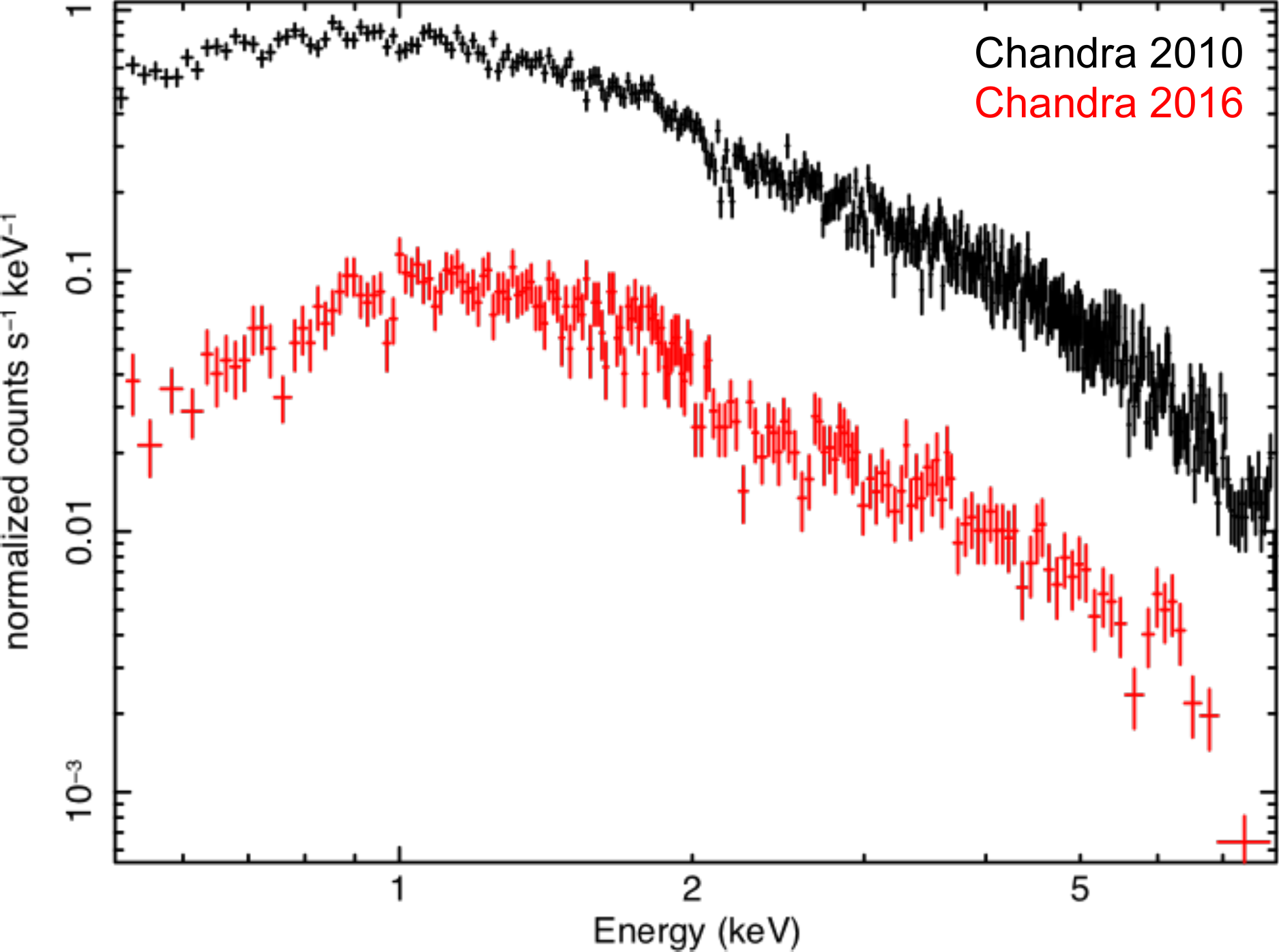}
  \includegraphics[scale=0.37]{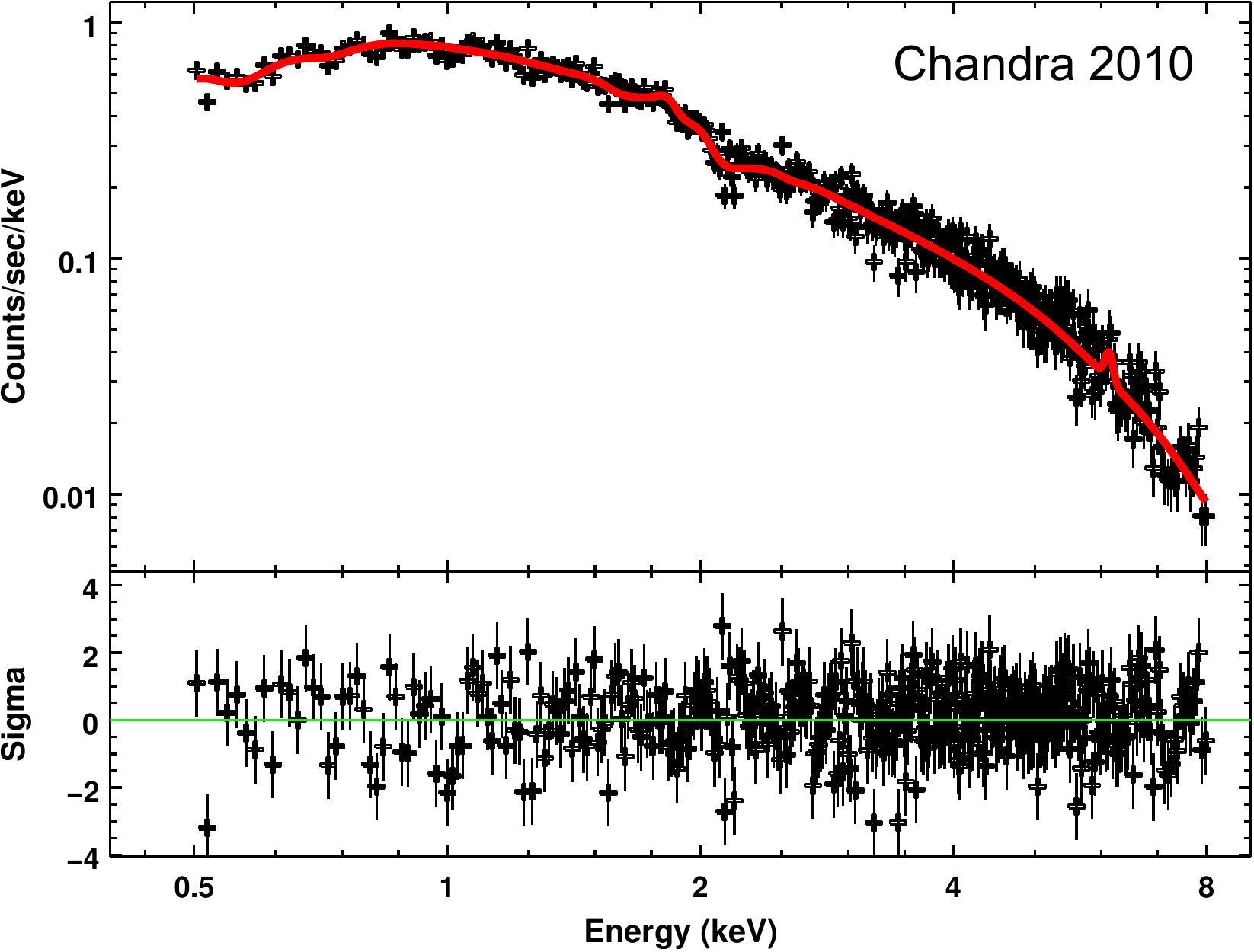}~
  \includegraphics[scale=0.35]{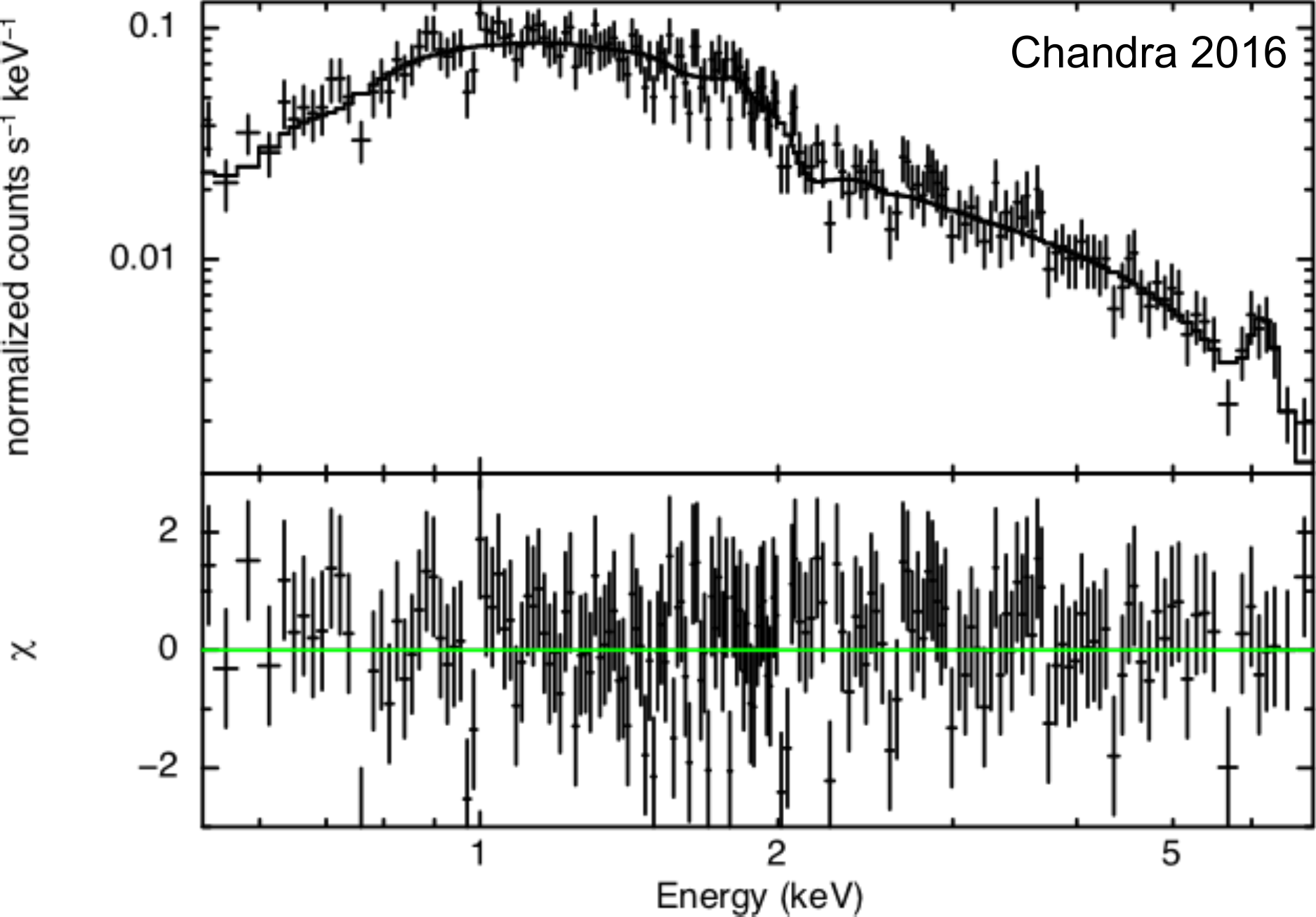}~
 \caption{\label{ch_spectra} {\it Top}: {\it Chandra} spectra of Mrk 1018 from 2010 ({\it black}) and 2016 ({\it red}). Between epochs, the aparent X-ray flux dimmed (overall normalization decreased) and Fe K$\alpha$ line emission (at 6.4 keV) becomes more pronounced: the Fe K$\alpha$ EW increased from 0.18$^{+0.17}_{-0.12}$ keV to 0.61$^{+0.27}_{-0.25}$ keV, based on our phenomenological model fits to the data. {\it Bottom left}: Powerlaw plus Gaussian plus pileup fit to the {\it Chandra} spectrum from 2010. {\it Bottom right}: Powerlaw plus Gaussian fit to the {\it Chandra} spectrum from 2016. While we fitted the {\it Chandra} spectrum from 2016 in \textsc{XSpec}, we fitted the spectrum from 2010 in \textsc{Sherpa} to account for pileup with the \textsc{jdpileup} model.}
\end{figure*}

In the {\it Chandra} spectrum from 2016, Fe K$\alpha$ emission at 6.4 keV (rest-frame) is visually apparent while it is not as discernable in the observation from 2010 (Figure \ref{ch_spectra}, top). However, we can test whether this feature may be present in the earlier epoch observation by modeling the {\it Chandra} spectrum phenomenologically. In \textsc{SHERPA}, we modeled the spectrum with a powerlaw to account for the AGN continuum, added a Gaussian component at the  energy of the Fe K$\alpha$ emission line, and included the \textsc{jdpileup} model to correct for the effects of pile-up.\footnote{The only free parameters in the \textsc{jdpileup} model are $\alpha$, which parameterizes the grade migration and gives the probability that a piled event is not rejected as ``bad'' by the spacecraft software, and $f$, the fraction of events to which the pile-up model will be applied. See http://cxc.harvard.edu/ciao/download/doc/pileup\_abc.pdf for more information.} From this exercise, we find an Fe K$\alpha$ flux of 2.8$^{+2.6}_{-1.9} \times 10^{-5}$ photons s$^{-1}$ cm$^{-2}$, with an equivalent width (EW) of 0.18$^{+0.17}_{-0.12}$ keV.

Is the Fe K$\alpha$ flux consistent with that from the 2016 observation, given the effects of pileup on the former observation? To test this, we fitted the {\it Chandra} spectrum from 2016 with the same phenomenological model as above, sans the pileup model. We note that the Fe K$\alpha$ EW significantly increased during this second epoch observation (0.61$^{+0.27}_{-0.25}$ keV). Though the line flux is somewhat lower (8.9$^{+3.9}_{-3.6} \times 10^{-6}$ photons s$^{-1}$ cm$^{-2}$), it is consistent within the error bars with that measured in 2010. 

We then set up a toy model to describe the {\it Chandra} spectrum from 2010 assuming the same Fe K$\alpha$ line strength observed in 2016. We used the best fit powerlaw parameters returned from fitting the 2010 spectrum to describe the AGN continuum in its bright state, plus the Gaussian parameters from fitting the 2016 spectrum to mimic the Fe K$\alpha$ emission observed in the faint state (see Table \ref{empir_fits}). We ran {\it Chandra} ray trace simulations with \textsc{marx} \citep{marx} to produce a simulated {\it Chandra} ACIS-S image of this model description. Using \textsc{marxpileup}, we simulated an events file that accounts for the effects of pileup and extracted a spectrum from this simulated image.

We fitted this simulated spectrum with the same phenomenological model as we used when fitting the spectrum from 2010 (i.e., \textsc{powerlaw} + \textsc{gauss} + \textsc{jdpileup}). We found an Fe K$\alpha$ flux of 1.8$^{+2.1}_{-1.4} \times 10^{-5}$ photons s$^{-1}$ cm$^{-2}$, consistent with the flux we measured from the earlier epoch {\it Chandra} spectrum (as well as the latter epoch {\it Chandra} spectrum), and an EW more akin to that from 2010 instead of 2016.

We thus conclude that the enhanced EW from the observation in 2016 is not due to an increase in the Fe K$\alpha$ flux but rather a decrease in the observed AGN continuum. This effect can be caused by 1) the AGN flux dimming or 2) an occulting cloud with sufficient column density ($N_{\rm H} > 10^{22}$ cm$^{-2}$) to block a significant portion of the transmitted light.

\begin{deluxetable*}{llll}[l]
  \tablewidth{0pt}
  \tablecaption{\label{empir_fits} Phenomenological Fit Parameters to {\it Chandra} Spectra of Mrk 1018}
  \tablehead{\colhead{Parameter} & \colhead{2010 Observation\tablenotemark{1}} &\colhead{2016 Observation} & \colhead{Simulated Spectrum\tablenotemark{2}}}
  \startdata
  $\Gamma$                                            & 1.97$^{+0.03}_{-0.04}$ & 1.66$^{+0.06}_{-0.05}$ & 1.92$^{+0.05}_{-0.05}$  \\
  Powerlaw Normalization (10$^{-3}$)\tablenotemark{3} & 5.4$^{+0.4}_{-0.5}$    & 0.30$^{+0.02}_{-0.01}$  & 4.2$^{+2.2}_{-1.0}$    \\
  Gauss $E$\tablenotemark{4}                          & 6.37$^{+0.05}_{-0.06}$ & 6.42$^{+0.11}_{-0.10}$  & 6.30$^{+0.14}_{-0.14}$ \\
  Gauss $\sigma$                                      & 0.01\tablenotemark{5} & 0.16$^{+0.10}_{-0.08}$  & $<$0.13              \\
  Gauss normalization(10$^{-5}$)\tablenotemark{6}      & 2.8$^{+2.6}_{-1.9}$   & 0.89$^{+0.39}_{-0.37}$   & 1.8$^{+2.1}_{-1.4}$   \\
  Fe K$\alpha$ Flux (10$^{-13}$ erg s$^{-1}$ cm$^{-2}$)  & 2.6$^{+2.4}_{-1.8}$    & 0.86$^{+0.38}_{-0.36}$  & 1.7$^{+2.0}_{-1.3}$    \\
  Fe K$\alpha$ EW (keV)                               & 0.18$^{+0.17}_{-0.12}$ & 0.61$^{+0.27}_{-0.25}$  & 0.12$^{+0.14}_{-0.09}$  

  \enddata
  \tablenotetext{1}{This model also included a \textsc{jdpileup} component, where we found a pile-up fraction of 30\%, $\alpha$=0.56$^{+0.13}_{-0.02}$, and $f$=0.927$^{+0.010}_{-0.007}$.}
  \tablenotetext{2}{Our toy model where the AGN continuum is described by the powerlaw fit parameters from the bright state and the Fe K$\alpha$ line emission is set by the Gaussian fit parameters from the dim state. We fitted this simulated spectrum with a \textsc{powerlaw + Gaussian + pileup model}.}
  \tablenotetext{3}{Powerlaw normalization in units of photons keV$^{-1}$ cm$^{-2}$ s$^{-1}$ at 1 keV.}
  \tablenotetext{4}{Rest-frame energy.}
  \tablenotetext{5}{Frozen to a lower limit of 0.01 keV, which is several times lower than the {\it Chandra} spectral resolution.}
  \tablenotetext{6}{Total photons cm$^{-2}$ s$^{-1}$ in line.}
\end{deluxetable*}

\subsection{Increase in Obscuration or Decrease in AGN Power?}
From our phenomenological modeling, we find that the hard X-ray (2-10 keV) flux decreased by an order of magnitude between epochs, from 1.46$^{+0.10}_{-0.13} \times 10^{-11}$ erg s$^{-1}$ cm$^{-2}$ to 1.31$^{+0.09}_{-0.04} \times 10^{-12}$ erg s$^{-1}$ cm$^{-2}$, which is a more significant drop than that found by \citet{husemann}. They found a lower hard X-ray flux from the bright state spectrum than what we do (9.2$\pm 0.2 \times 10^{-12}$ erg s$^{-1}$ cm$^{-2}$), likely due to the different methods employed to deal with pile-up on the {\it Chandra} spectrum. They thus find a lower decline in the hard X-ray flux (factor of 7.5).

To gain insight as to whether the drop in X-ray flux is due to obscuration, we fit the {\it Chandra} spectrum from 2016 with the {\it NuSTAR} FPMA and FPMB spectra in \textsc{XSpec} using the physically motivated MYTorus model which self-consistently fits the trasnsmitted AGN contintuum, reflected emission, and fluoresent Fe K$\alpha$ and Fe K$\beta$ emisison lines. In the MYTorus model, the X-ray reprocessor takes the form of a uniform torus (in the default mode) with a fixed opening angle of 60$^{\circ}$. With an inclination angle greater than 60$^{\circ}$, the line-of-sight intersects the torus. An inclination angle of 0$^{\circ}$ is face-on, such that there is no absorption along the line-of-sight, yet reprocessing still occurs in the circumnuclear medium, shaping the X-ray spectrum we observe. A schematic of this model is:
\begin{equation}
  \begin{split}
  model = MYTorusZ \times powerlaw \, + \\
  A_{\rm s} \times (MYTorusS + MYTorusL),
  \end{split}
\end{equation}
where the powerlaw describes the intrinsic AGN emission, \textsc{MYTorusZ} represents the line-of-sight attenuation, \textsc{MYTorusS} describes the Compton-scattered emission, \textsc{MYTorusL} accounts for the fluorescent line emission,\footnote{Here we assume the Fe K$\alpha$ line is not resolved.} and $A_{\rm S}$ is the relative normalization between the Compton-scattered emission with respect to the transmitted continuum. This normalization constant encompasses several unknown quantities, including time delays between the transmitted continuum and scattered and line emission, different torus half opening angles and/or elemental abundances than those assumed in the MYTorus model set up. We include a constant factor in the model to account for cross-calibration normalizations between {\it Chandra} and {\it NuSTAR}.

\begin{figure*}
  \centering
  \includegraphics[scale=0.40]{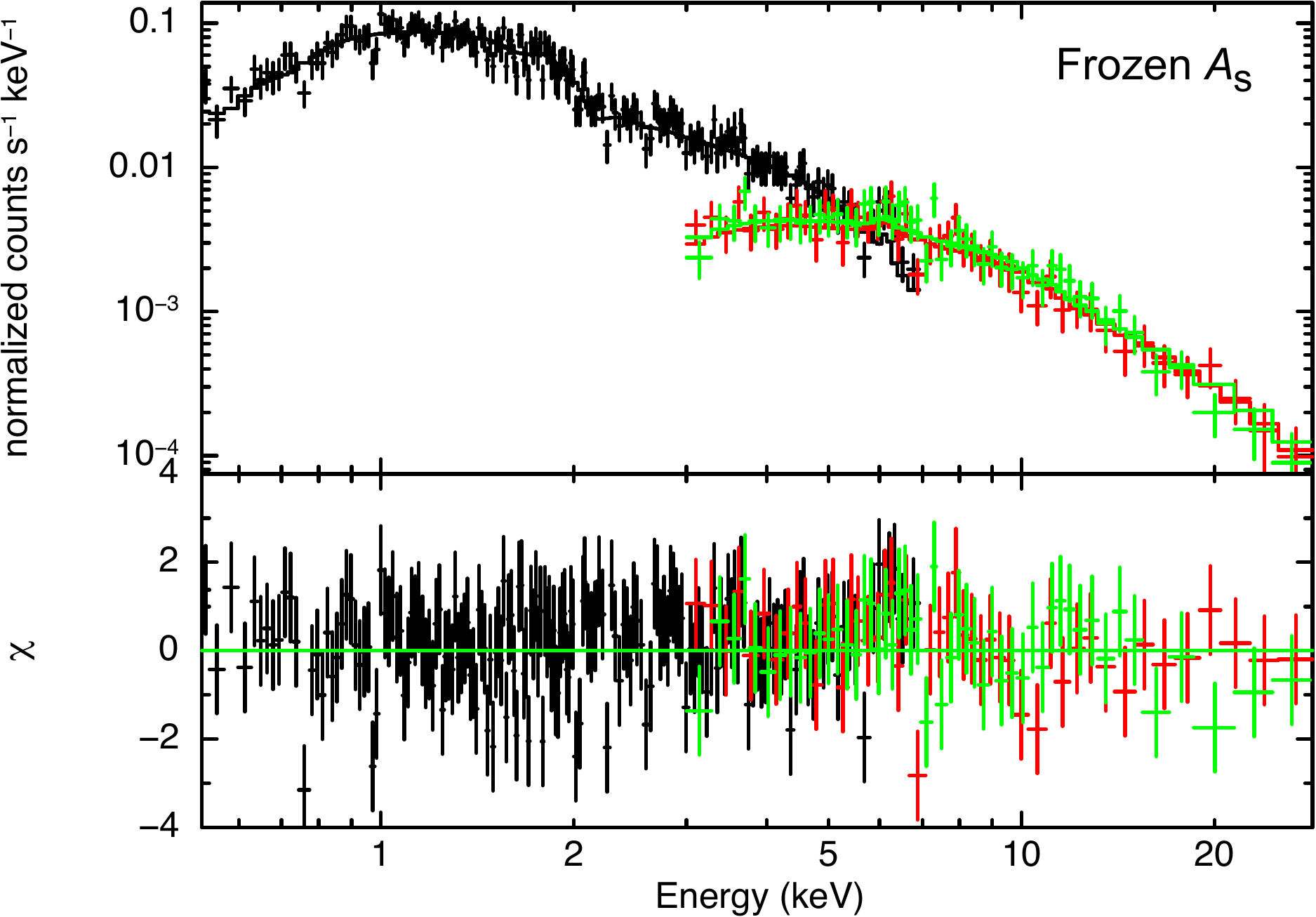}~
  \includegraphics[scale=0.40]{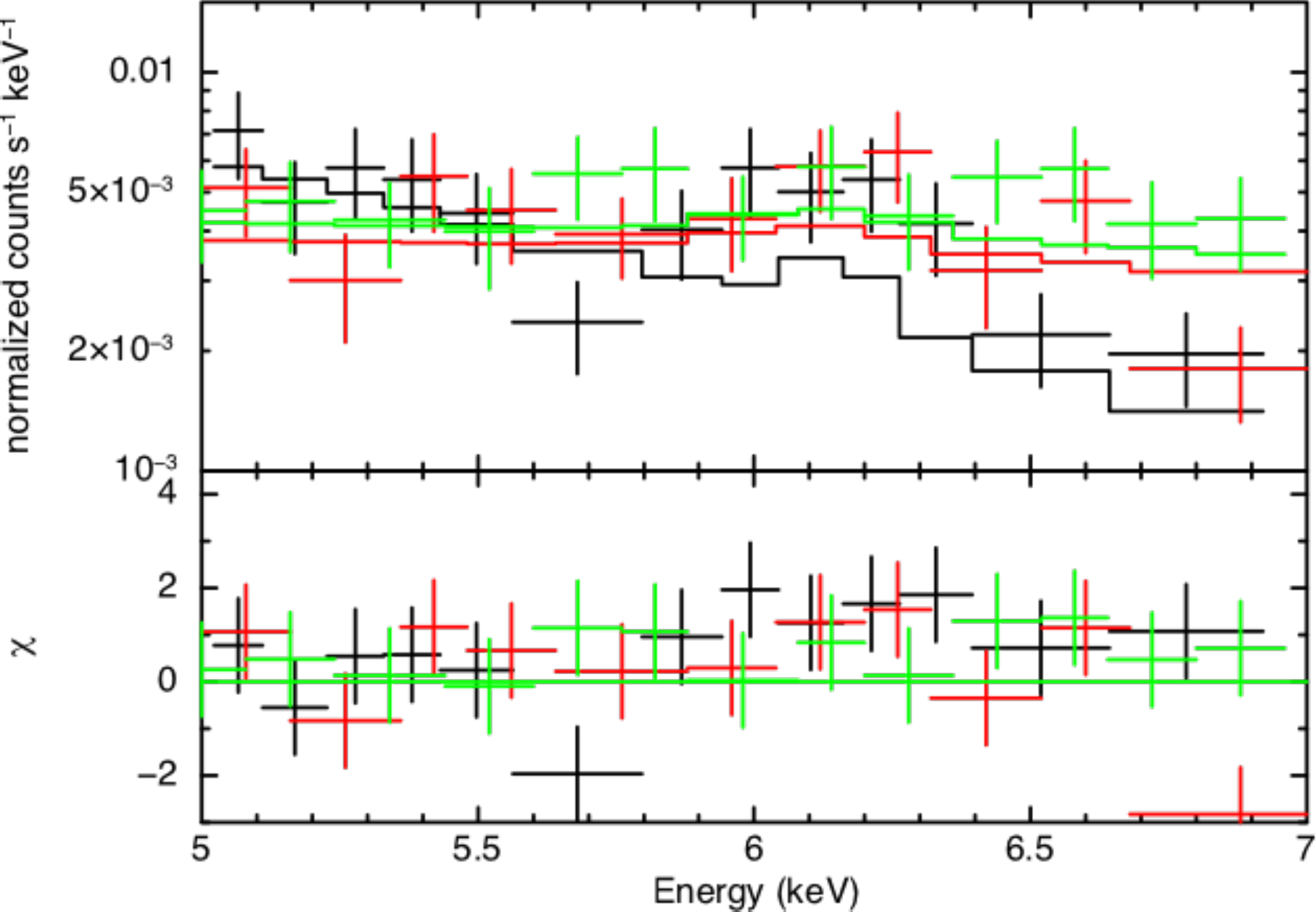}
 \caption{\label{frozeAs_figs} {\it Chandra} (black), {\it NuSTAR} FPMA (red), and {\it NuSTAR} FPMB (green) spectra fit with MYTorus, where $A_{\rm S}$, the relative normalization between the transmitted and Compton scattered emission, is frozen to unity. Though this realization of the model decently reproduces the global spectral shape ({\it left}), it is unable to accommodate the Fe K$\alpha$ emission ({\it right}).}
\end{figure*}

We begin by keeping $A_{\rm S}$ frozen to the default value of unity as this is often an adequate description of X-ray spectra of AGN \citep{yaqoob,lamassa2014}. We find that the inclination angle is constrained to be under 60$^{\circ}$, indicating that our line-of-sight does not intercept the torus. As the exact value of the inclination angle is otherwise unconstrained, we freeze it to 0$^{\circ}$, consistent with a face-on geometry. As Figure \ref{frozeAs_figs} depicts, we obtain a good fit to the global spectrum ($\chi^2$=245.4 for 271 degrees of freedom), but fail to accurately model the Fe K$\alpha$ line. 

\begin{figure*}
  \centering
  \includegraphics[scale=0.40]{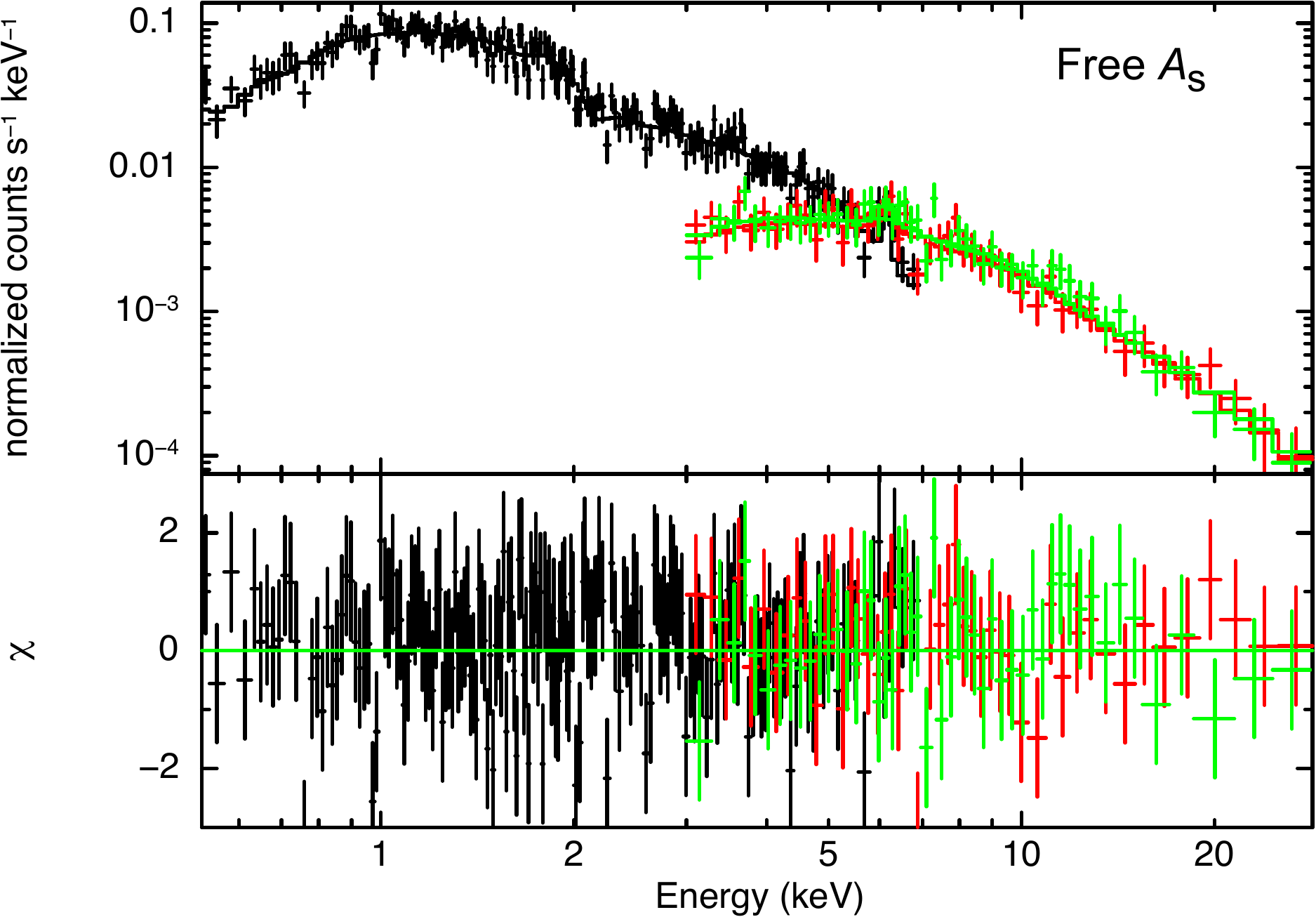}~
  \includegraphics[scale=0.40]{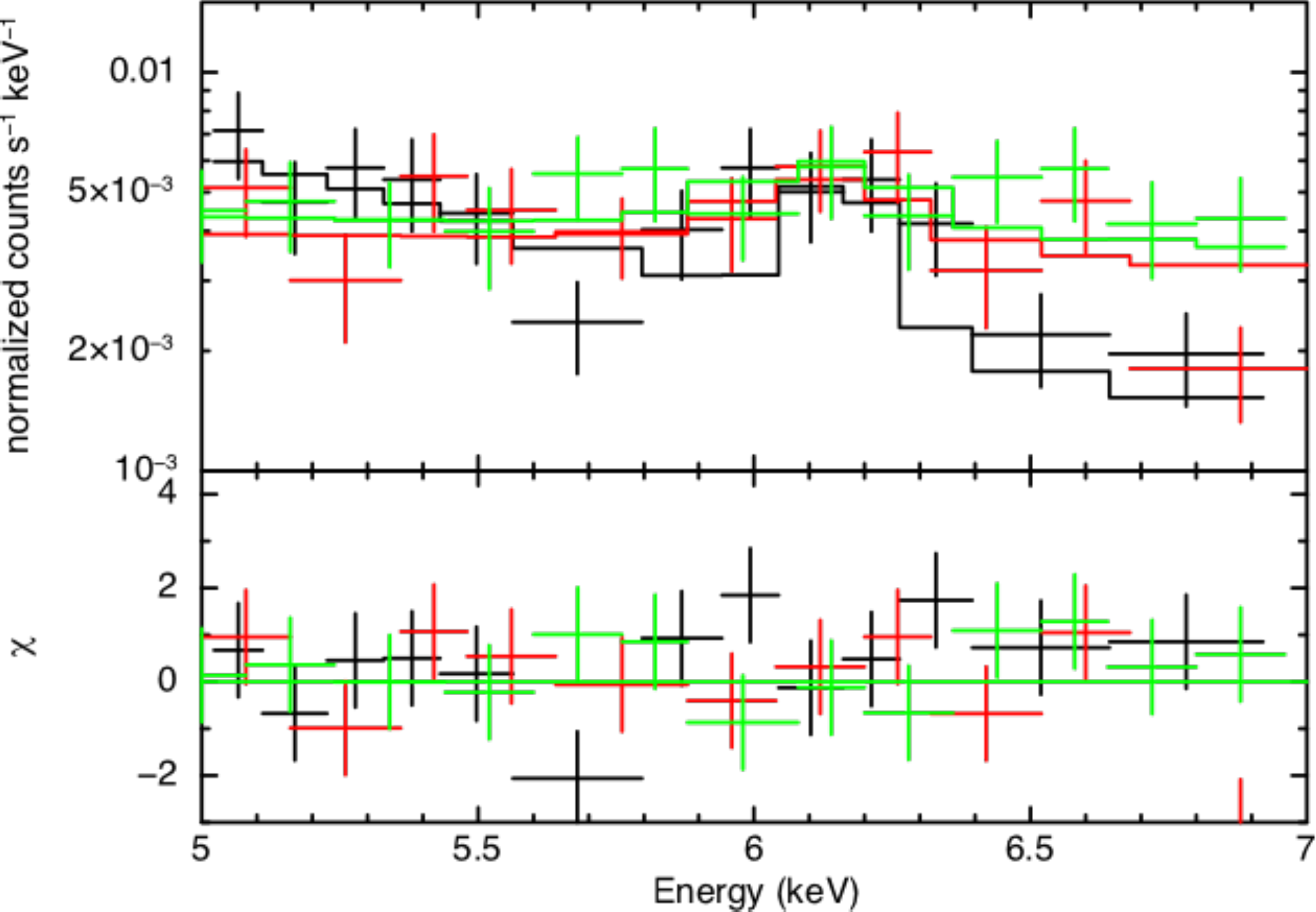}
 \caption{\label{freeAs_figs} MYTorus fit to the X-ray spectra of Mrk 1018 with $A_{\rm S}$ as a free paramater. Both the global spectral shape ({\it left}) and the Fe K$\alpha$ complex ({\it right}) are well modeled. Here, we find no absorption along the line-of-sight, i.e., the torus is consistent with a face-on geometry and has a column density of 5.38$^{+14}_{-4.0} \times 10 ^{22}$ cm$^{-2}$. This gas reprocesses the X-ray photons and produces the fluorescent line emission.}
\end{figure*}

We then allow the relative normalization between the transmitted and Compton scattered components to be a free parameter. As shown in Figure \ref{freeAs_figs}, the Fe K$\alpha$ emission line is now well modeled, and according to the $f$-test, the improvement to the fit is statistically significant ($P = 0.0001$). Again, we find that the inclination angle is below 60$^{\circ}$, indicating that the gas reprocessing the X-ray emission does not intersect our line-of-sight. This result confirms the findings of \citet{husemann} who found no line-of-sight absorption. However, the global medium, out of the line-of-sight, has a measureable, non-zero column density of 5.38$^{+14}_{-4.0} \times 10^{22}$ cm$^{-2}$.  We note that the exact value of the column density depends also on the iron abundance, though as MYTorus assumes a fixed iron abundance  of solar, we are unable to test the effects of varying the iron abundance on $N_{\rm H}$.

\begin{figure}
  \centering
  \includegraphics[scale=0.35,angle=270]{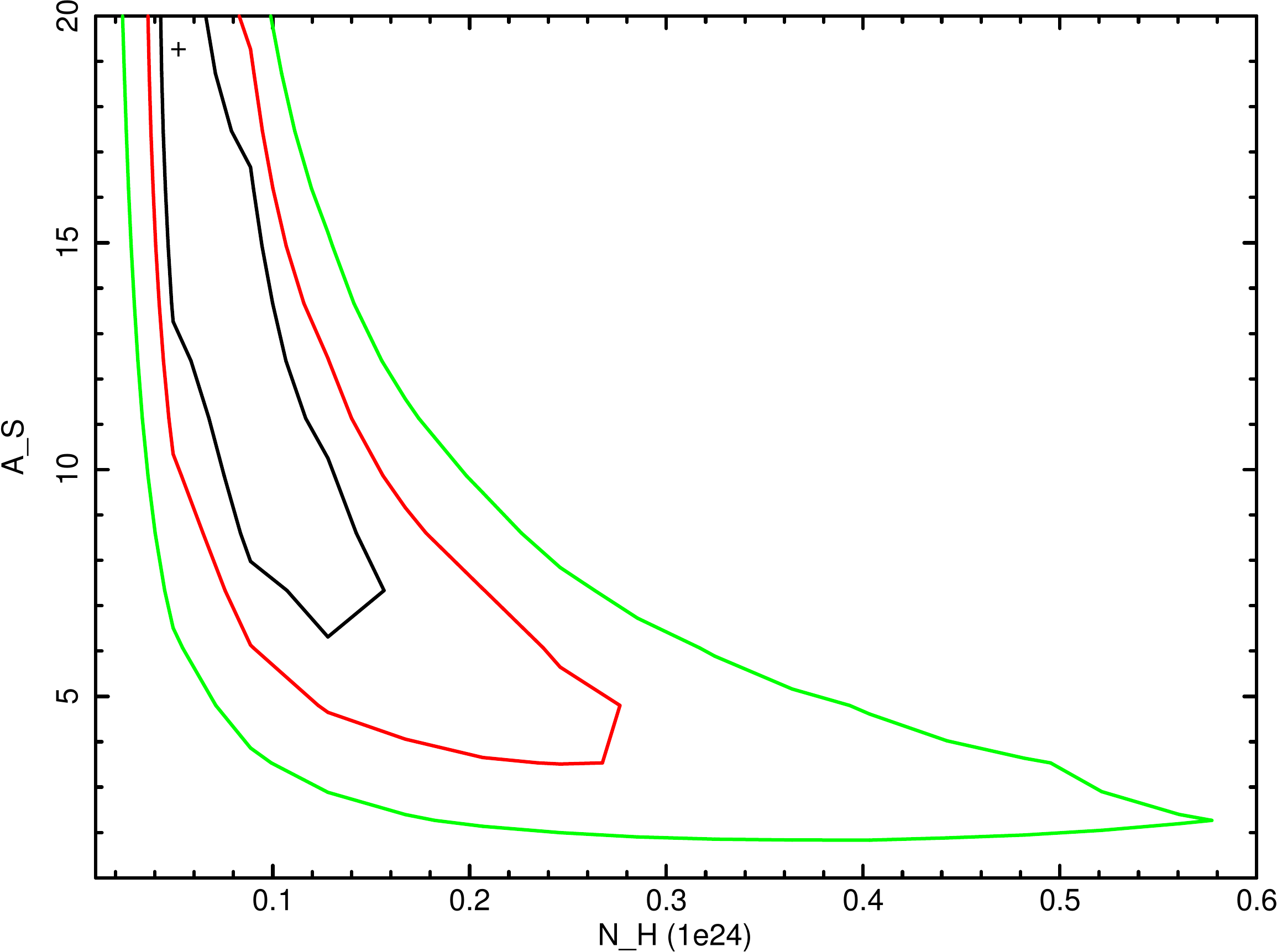}
 \caption{\label{con_as_nhs} Contour plot of the column density ($N_{\rm H}$) and relative normalization between the transmitted continuum and Compton-scattered emission ($A_{\rm S}$), with contours at the 68\% ({\it black}), 90\% ({\it red}), and 99\% ({\it green}) confidence levels. The remarkably high normalization ($>3$ at the 99\% confidence level) indicates time delays between the continuum and Compton-scatterred emission.}
\end{figure}

In order to achieve this acceptable fit to both the global spectrum and the Fe K$\alpha$ line, the relative normalization of the Compton scattered emission was forced to a remarkably high value.  In Figure \ref{con_as_nhs}, we show the two-parameter confidence contours of $A_{S}$ and $N_{\rm H}$, where $A_{\rm S}$ is constrained to be over 3 at the 99\% confidence level. Though $A_{\rm S}$ encapsulates our ignorance of several unknown parameters, time delays between the continuum and the scattering medium would play the biggest role in boosting $A_{\rm S}$ to such a high level.

Finally, we test whether we can measure an upper limit to the column density along the line-of-sight. To measure a line-of-sight $N_{\rm H}$ that is independent from the circumnuclear gas out of the line-of-sight which produces the Fe K$\alpha$ emission, we run MYTorus in ``decoupled mode.'' Here, the \textsc{MYTorusS} and \textsc{MYTorusL} components have an inclination angle fixed at 0$^{\circ}$, mimicking a face-on torus. The transmitted continuum that is potentially attenuated by line-of-sight absorption is generally represented by \textsc{MYTorusZ} in the decoupled realization of the model. However, since this model component has a lower limit of $N_{\rm H} = 10^{22}$ cm$^{-2}$, which we find to be too high for this source, we replace this component with \textsc{zphabs}; as Compton-scattering does not appreciably affect the spectrum at column densities below 10$^{22}$ cm$^{-2}$, this swap does not affect the self-consistency of the model. From this excercise, we find that the line-of-sight column density is below $3\times10^{20}$ cm$^{-2}$ at the 90\% confidence level, consistent with being  X-ray unabsorbed along the line-of-sight.

We thus conclude that the increase in the Fe K$\alpha$ EW between 2010 and 2016 is due to a precipitious drop in the AGN continuum. The gas responsible for reprocessing the X-ray emission has not yet fully responded to this change. However, \citet{mcelroy} demonstrated that the mid-infrared {\it WISE} flux of this source at 3.4 $\mu$m and 4.6$\mu$m, ascribed to torus emission, decreased between 2010 and 2015. Though the Fe K$\alpha$ line flux between 2010 and 2016 is consistent given the errors, the nominal flux is lower during the second epoch, hinting that the X-ray reprocessing gas is starting to respond to the change from the central engine, consistent with the mid-infrared variability. Additionally, the dusty torus and X-ray reprocessing gas can span different size scales, which leads to differential time lags with respect to the direct continuum.  If this interpretation is correct, we expect that the Fe K$\alpha$ EW will decrease over time. Continued monitoring of this source will reveal whether the Fe K$\alpha$ EW responds, and provide insight into the size of the reprocessing region and how it relates to the geometric structure of the dusty torus, based on the time delay of this response.

\begin{deluxetable}{ll}[l]
  \tablewidth{0pt}
  \tablecaption{\label{mytorus_params} MYTorus Fit Parameters}
  \tablehead{\colhead{Parameter} & \colhead{Value} }
  \startdata
  $\Gamma$                                            & 1.76$^{+0.06}_{-0.06}$ \\
  Powerlaw Normalization (10$^{-4}$)\tablenotemark{1} & 3.03$^{+0.13}_{-0.17}$ \\
  $N_{\rm H}$ (10$^{22}$ cm$^{-2}$)                     & 5.38$^{+14}_{-4.0}$  \\
  $\chi^2$ (DOF)                                      & 232.5 (270)
  \enddata
  \tablenotetext{1}{Normalization in units of photons keV$^{-1}$ cm$^{-2}$ s$^{-1}$ at 1 keV.}
\end{deluxetable}

\section{Conclusions}
We have undertaken a rigorous analysis of the {\it Chandra} and {\it NuSTAR} spectra of changing-look AGN Mrk 1018 to determine the cause of the reported decrease in X-ray flux \citep{husemann}. We modeled the effects of pile-up on the {\it Chandra} spectrum from 2010 to assess as accurately as possible the intrinsic AGN powerlaw continuum during the bright state of Mrk 1018. We used a phenomenological model to fit the {\it Chandra} spectra from 2010 and 2016, where we included a Gaussian component for Fe K$\alpha$ line emission. From this exercise, we demonstrated that the Fe K$\alpha$ EW increased from 0.18$^{+0.17}_{-0.12}$ keV to 0.61$^{+0.27}_{-0.25}$ keV between 2010 and 2016.

To test whether the Fe K$\alpha$ line flux was consistent between epochs, such that we can attribute the change in the EW to suppression of the continuum, we used the {\it Chandra} ray-tracing simulator \textsc{marx} to account for the effects of pile-up on the spectrum. Here, we set-up a toy model to describe the AGN spectrum from 2010, where the AGN continuum (i.e., powerlaw) represents that which we measured when fitting the spectrum from 2010 with the \textsc{jdpileup} model and the Fe K$\alpha$ line flux matches that which we measured from the {\it Chandra} spectrum in 2016. When simulating a source with this spectrum in \textsc{marx}, including the effects of pile-up, we find an Fe K$\alpha$ flux consistent with that which we measured from the {\it Chandra} 2010 spectrum. We thus conclude that the variation in the Fe K$\alpha$ EW is due to attentuation of the continuum and not an increase in the line flux.

We jointly fitted the {\it Chandra} spectrum from 2016 with the {\it NuSTAR} spectra from 2016 using the MYTorus model \citep{mytorus}, which self-consistently models the transmitted, Compton scattered, and fluorescent line emission in obscured AGN. This model indicates that the orientation of the circumnuclear obscuration is consistent with a face-on geometry, with a column density of N$_{\rm H}$ = 5.38$^{+14}_{-4.0} \times 10^{22}$ cm$^{-2}$, and no absorption is present along the line-of-sight. The X-ray reprocessing gas produces Fe K$\alpha$ line emission which is subsequently reflected into our line-of-sight. In order to fit both the global spectral shape and the Fe K$\alpha$ line, the relative normalization between the transmitted continuum and Compton scattered emission is forced to extremely high values.

This result can be understood via time lags between the continuum and the reprocessing gas: the torus has not yet fully responded to the decrease in flux from the AGN continuum. Though mid-infrared emission from this source shows a decline \citep{mcelroy}, consistent with the dusty torus starting to respond to the fading of the AGN, the Fe K$\alpha$ line, formed within the X-ray reprocessing gas, still has to catch up to this change, which would manifest as a decrease in the Fe K$\alpha$ EW. The timescale of this change would indicate the distance to the X-ray reprocessing gas and how it relates to the dusty torus.

\acknowledgements
We thank the anonymous referee for a careful reading of this manuscript. S.M.L. is grateful for helpful conversations with J. R. R. Rigby when preparing this manuscript.The scientific results reported in this article are based to a significant degree on observations made by the {\it Chandra} X-ray Observatory. This research has made use of software provided by the {\it Chandra} X-ray Center (CXC) in the application packages CIAO, ChIPS, and Sherpa.

{\it Facilities:} \facility{CXO},\facility{NuSTAR}

\end{document}